\documentstyle[11pt,newpasp,twoside,epsf]{article}
\def\edcomment#1{\iffalse\marginpar{\raggedright\sl#1\/}\else\relax\fi}
\reversemarginpar

\newcommand{\Lbol}{$L_{\rm bol}$}
\newcommand{\Ledd}{$L/L_{\rm Edd}$}
\newcommand{\Mbh}{$M_{\rm BH}$}
\newcommand{\Hb}{H$\rm\beta$}

\begin{document}
\title{What is the Origin of the Baldwin Effect in the BLR and the NLR?}
\author{Alexei Baskin \& Ari Laor}
\affil{Physics Department, Technion, Haifa 32000, Israel;
alexei@physics.technion.ac.il; laor@physics.technion.ac.il}

\begin{abstract}

The origin of the luminosity dependence of the strength of broad
and narrow emission lines in AGN (i.e. the Baldwin effect) is not
firmly established yet. We
explore this question using the Boroson \& Green sample of the
87 $z\le 0.5$ PG quasars. UV spectra of the C~IV region with a
sufficient S/N are available for 81 of the objects. We use these
spectra to explore the dependence of the C~IV EW on {\Mbh} and {\Ledd},
as deduced from optical spectra of the {\Hb} region. We find a strong correlation
of the C~IV EW with {\Ledd}. This may be the primary
correlation which drives the Baldwin effect for C~IV. A similar
correlation analysis using the C~IV FWHM instead of the {\Hb}
FWHM, yields an insignificant correlation. This suggests that the
C~IV FWHM provides a significantly less accurate estimate of
{\Mbh} compared to the {\Hb} FWHM.
We also measured the [O~III]~$\lambda 4363$/[O ~III]~$\lambda 5007$
line ratio for 72 objects in this sample, and use it to determine
whether density or covering factor control the large range in
[O ~III]~$\lambda 5007$ strength in AGN.

\end{abstract}

\section{Introduction}

The inverse correlation of emission line equivalent width (EW)
with luminosity, discovered by Baldwin (1977) for C~IV, was
intensively explored over the past 20 years (see a comprehensive
review by Osmer \& Shields 1999; and more recent studies by Green
et al. 2001; Croom et al. 2002; Dietrich et al. 2002; Richards et
al. 2002; and Shang et al. 2003). The physical origin for this
effect is not clearly established, but plausible explanations
include changes with luminosity of the ionizing continuum shape,
of the broad line region (BLR) covering factor, of the mean BLR
ionization level, or
possibly inclination effects. Recent studies established that
reasonably accurate estimates of the black hole mass ({\Mbh}) can be
obtained in AGN based on the continuum luminosity and
{\Hb} FWHM (e.g. Laor 1998). This opens up the possibility to explore whether
the Baldwin effect is related to {\Mbh}, and {\Ledd}. We
address this question in this paper, together with a related side
question on whether the C~IV FWHM can be used instead of the
{\Hb} FWHM as a comparably accurate measure of {\Mbh}.

The narrow forbidden lines in AGN show a larger range in EW
compared to the broad lines, and they also show a Baldwin
effect. For example, the [O~III]$\lambda 5007$ EW
ranges from $>150$~\AA\ to undetectable ($<0.5$~\AA) in the Boroson \&
Green (1992) sample, while the C~IV EW ranges from $\sim 5$~\AA, to $\sim
300$~\AA\ in the same sample (see below). What physical parameter
controls the [O~III] EW? Is it mostly density or is it the covering factor
of the illuminated narrow line region (NLR) gas?
If  it is controlled by covering factor variations, then the
[O~III]~$\lambda 4363$/[O~III]~$\lambda 5007$ ratio should be
independent of the [O~III]~$\lambda 5007$ EW.
However, if it is controlled by density variations in the NLR,
then the [O~III]~$\lambda 4363$/[O~III]~$\lambda 5007$ ratio should
increase with decreasing [O~III]~$\lambda 5007$ EW. We explore these
two possibilities below.

\section{The C~IV EW}

To explore the relation of the Baldwin effect with {\Mbh}
and \Ledd\ one needs high quality spectra of both the {\Hb} and
the C~IV regions. We use the Boroson \& Green (1992) optical
spectra of a complete and well defined sample of 87 $z\le 0.5$ AGN
from the BQS sample. This data set provides us with uniform and
systematic estimates of {\Mbh}, and {\Ledd}.
Archival UV spectra are available for 85 of the objects (47 HST,
38 IUE), of which 81 have sufficient S/N to measure the C~IV
EW, which we measured by fitting each
spectrum with 3 gaussian components in the range of $1470$~\AA\ to
$1620$~\AA. The {\Hb} FWHM and {\Lbol} were used for estimating
{\Mbh} and {\Ledd} for each object (e.g. Laor 1998).

The relations of the C~IV EW with {\Lbol} and {\Ledd}, together with
the corresponding Spearman rank-order correlation coefficients,
are plotted in Fig. 1. The C~IV EW has a significantly
tighter relation with {\Ledd} than with {\Lbol} (Rs increases from
$-0.154$ to $-0.592$). The correlation
with {\Mbh} is low (Rs=0.222, Pr=0.047).
We try to further reduce the scatter by including a third parameter
describing the optical emission from Boroson \& Green.
The parameter which most significantly improves the correlation
with {\Ledd} is the [OIII] EW (Fig. 2, left panel), which
yields Rs=-0.733. Other parameters which produce a comparably significant
improvement are $\alpha_{\rm ox}$ and the FeII/{\Hb} flux ratio.

Recent studies have used the C~IV FWHM, instead of the
{\Hb} FWHM, to estimate {\Mbh} and {\Ledd} (Vestergaard 2002;
Warner et al. 2003). Since the {\Hb} FWHM and C~IV FWHM are
rather poorly correlated, it is not clear whether the C~IV FWHM
is a useful substitute to the {\Hb} FWHM (when using {\Lbol} to estimate
the size of the C~IV emitting region). We repeated the correlation
analysis above using the C~IV FWHM for a revised {\Ledd} estimate
(Fig. 2, right panel). The revised {\Ledd} yields a low
correlation (Rs=-0.192), i.e. no significant improvement over the
correlation of the C~IV EW with {\Lbol}. This suggests that
the C~IV FWHM does not provide an {\Mbh} estimate of comparable
accuracy to the estimate based on the {\Hb} FWHM.

\begin{figure}
\plottwo{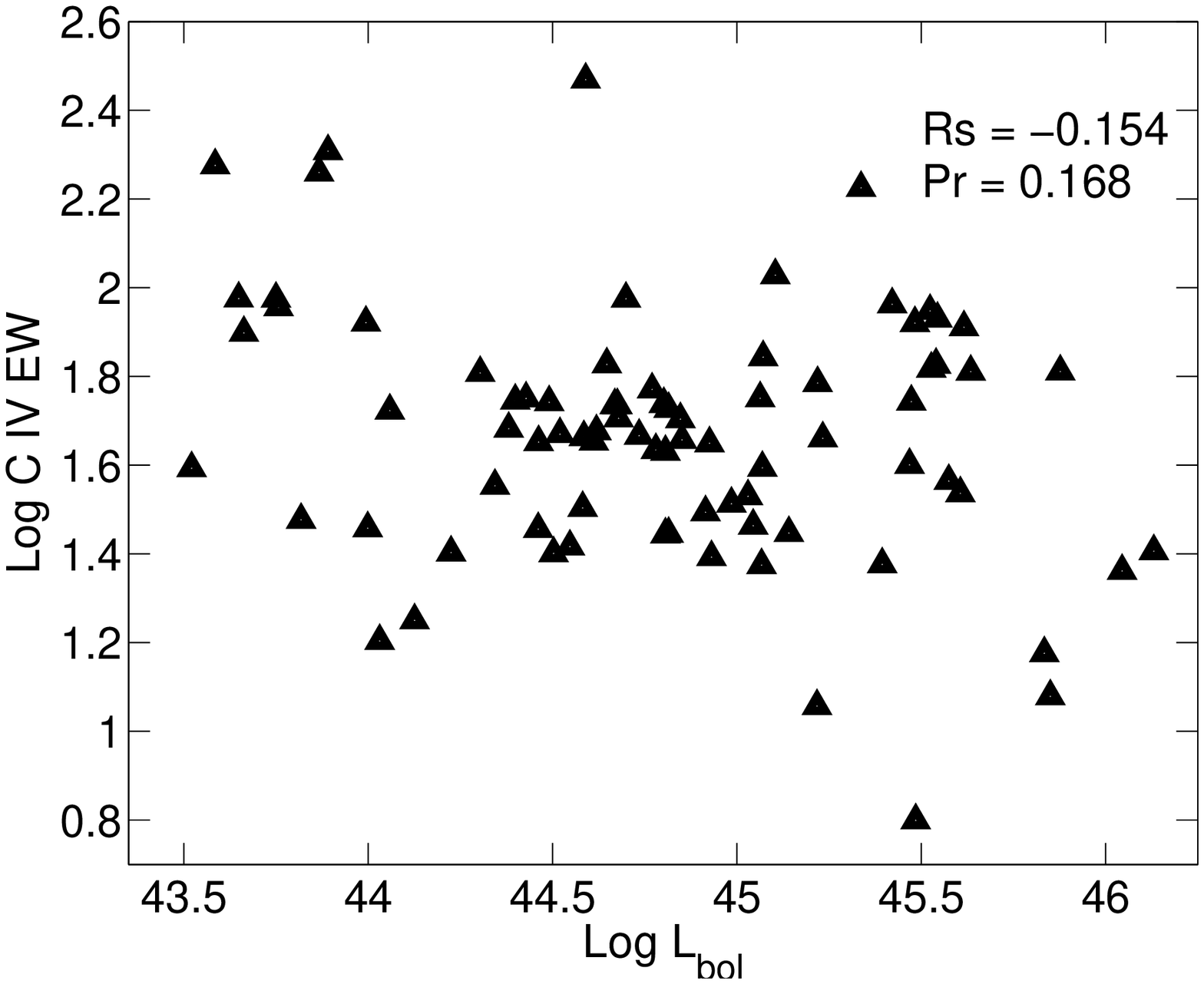}{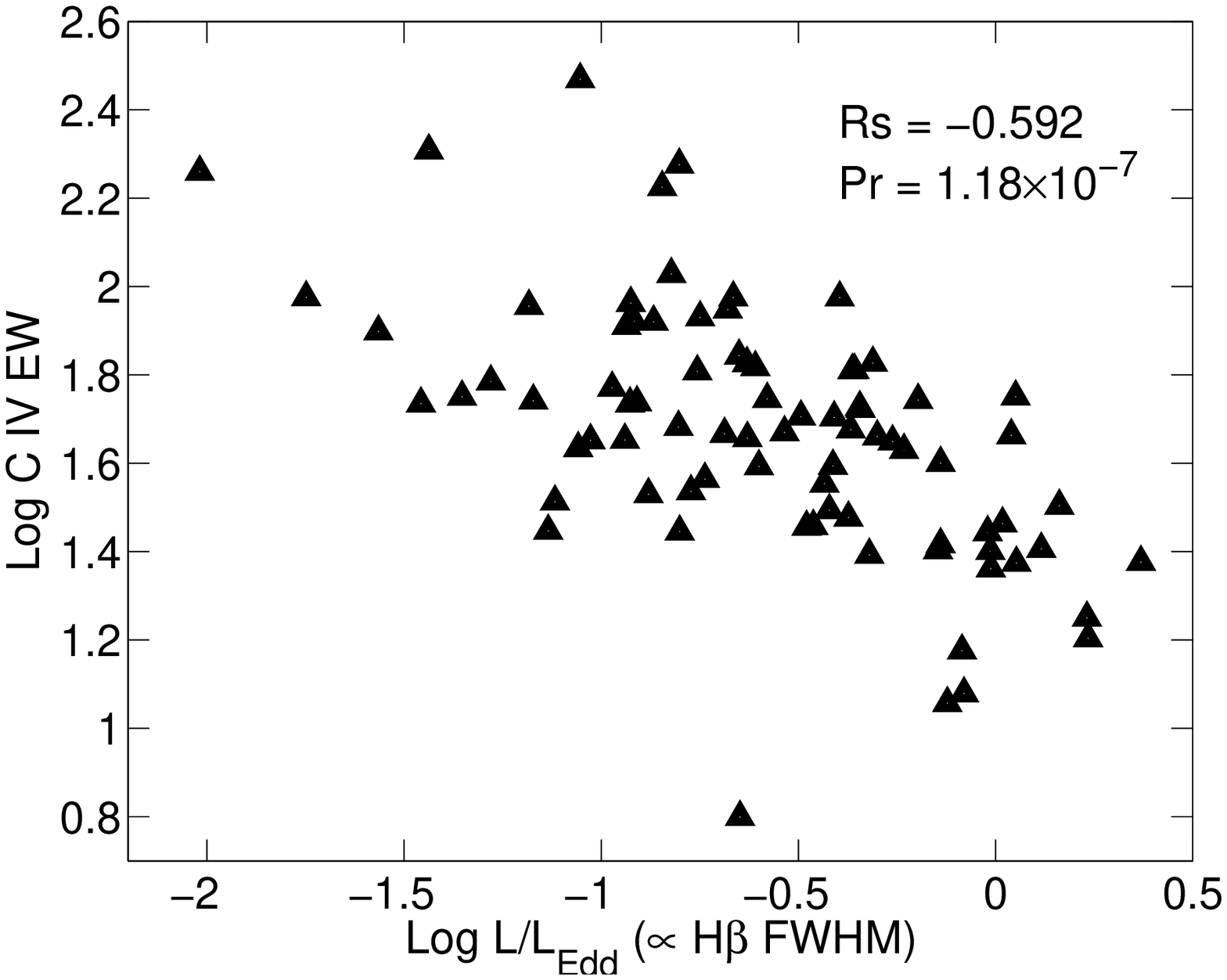}
\caption{Left: the Baldwin relation for the 81 BQS quasars.
Right: the dependence of C~IV EW on {\Ledd}.}
\end{figure}
\begin{figure}
\plottwo{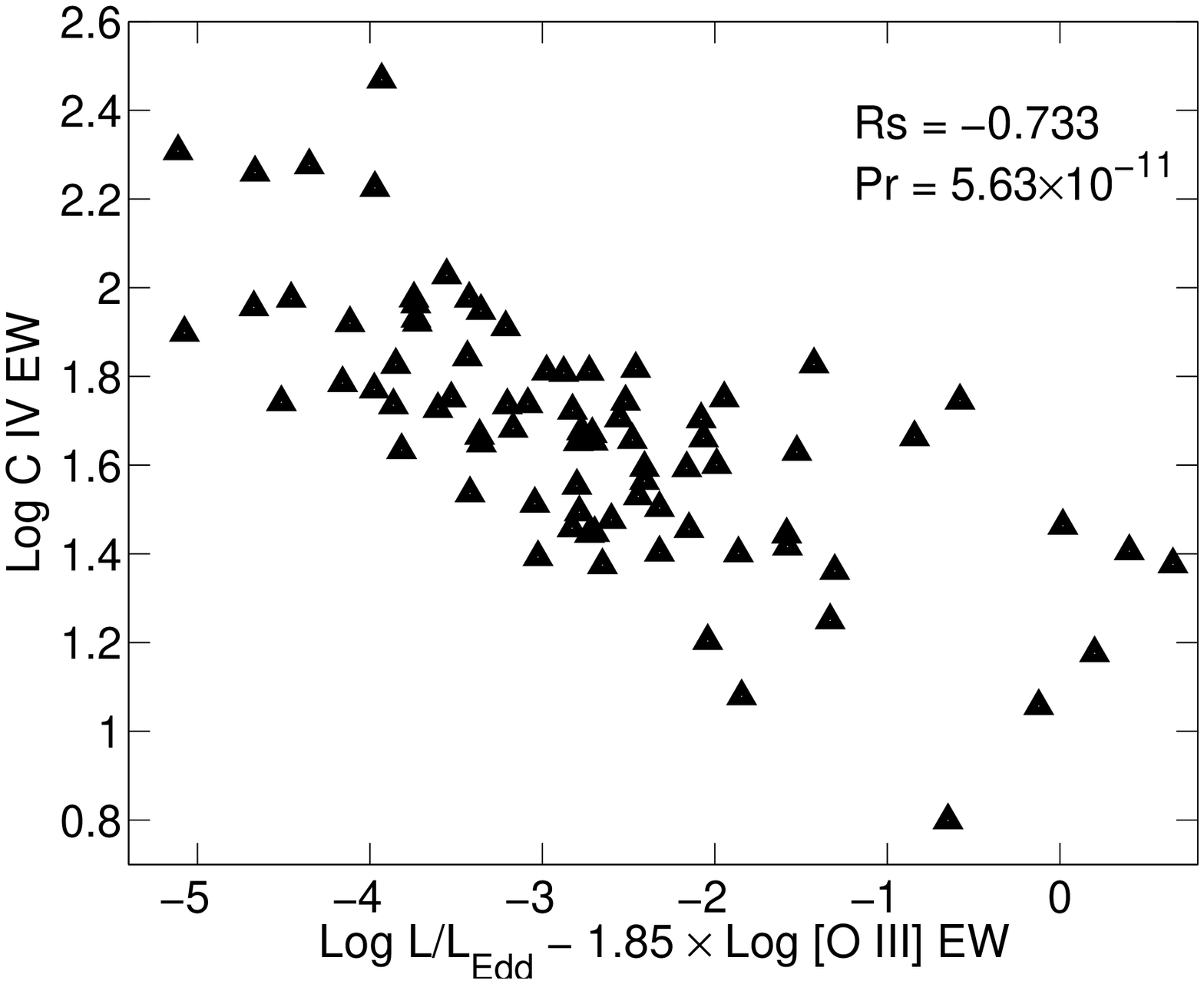}{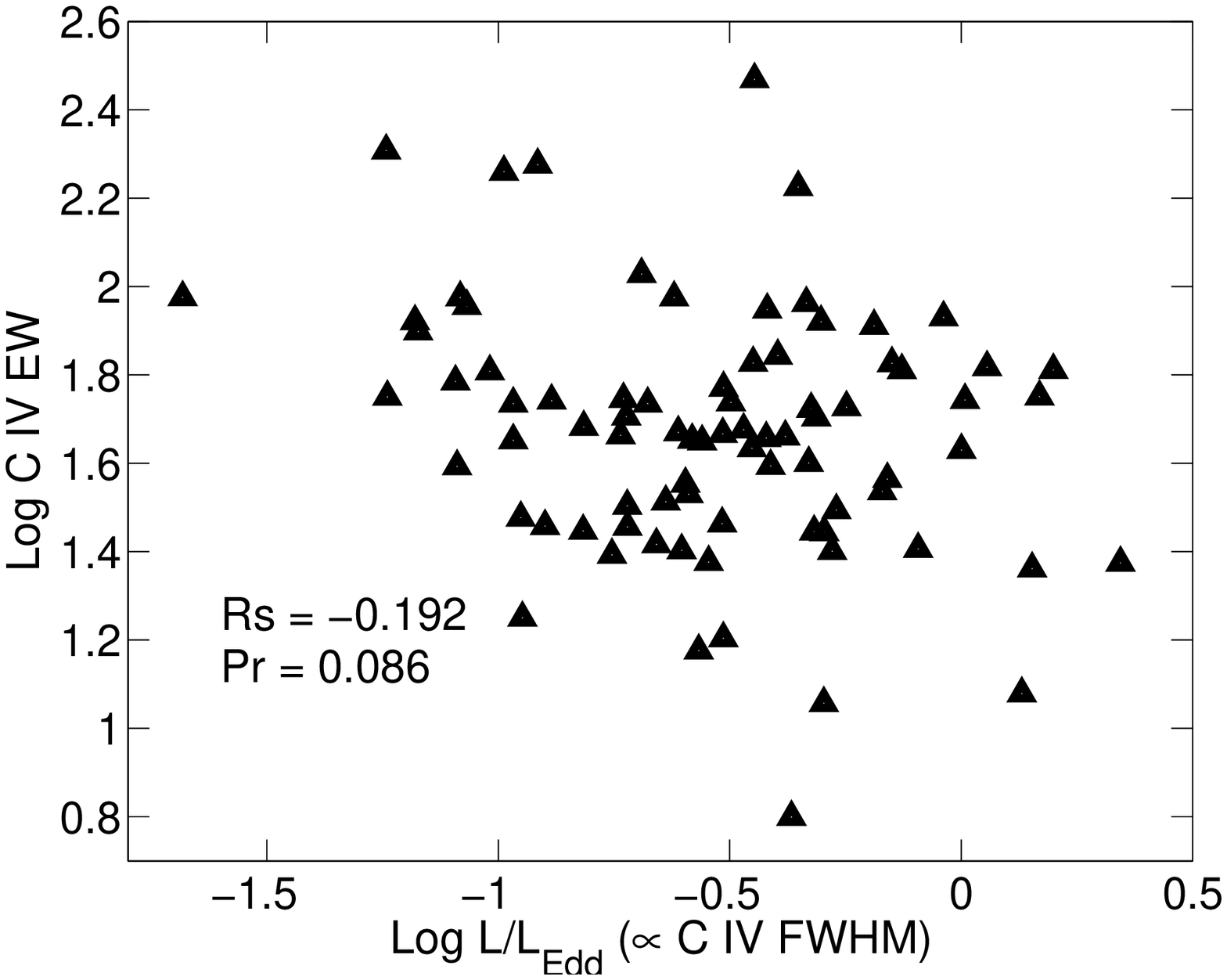}
\caption{Left: the scatter is further reduced when the [O~III] EW is included.
Right: the C~IV EW relation with {\Ledd} which is estimated using the C~IV FWHM.}
\end{figure}

\begin{figure}
\plottwo{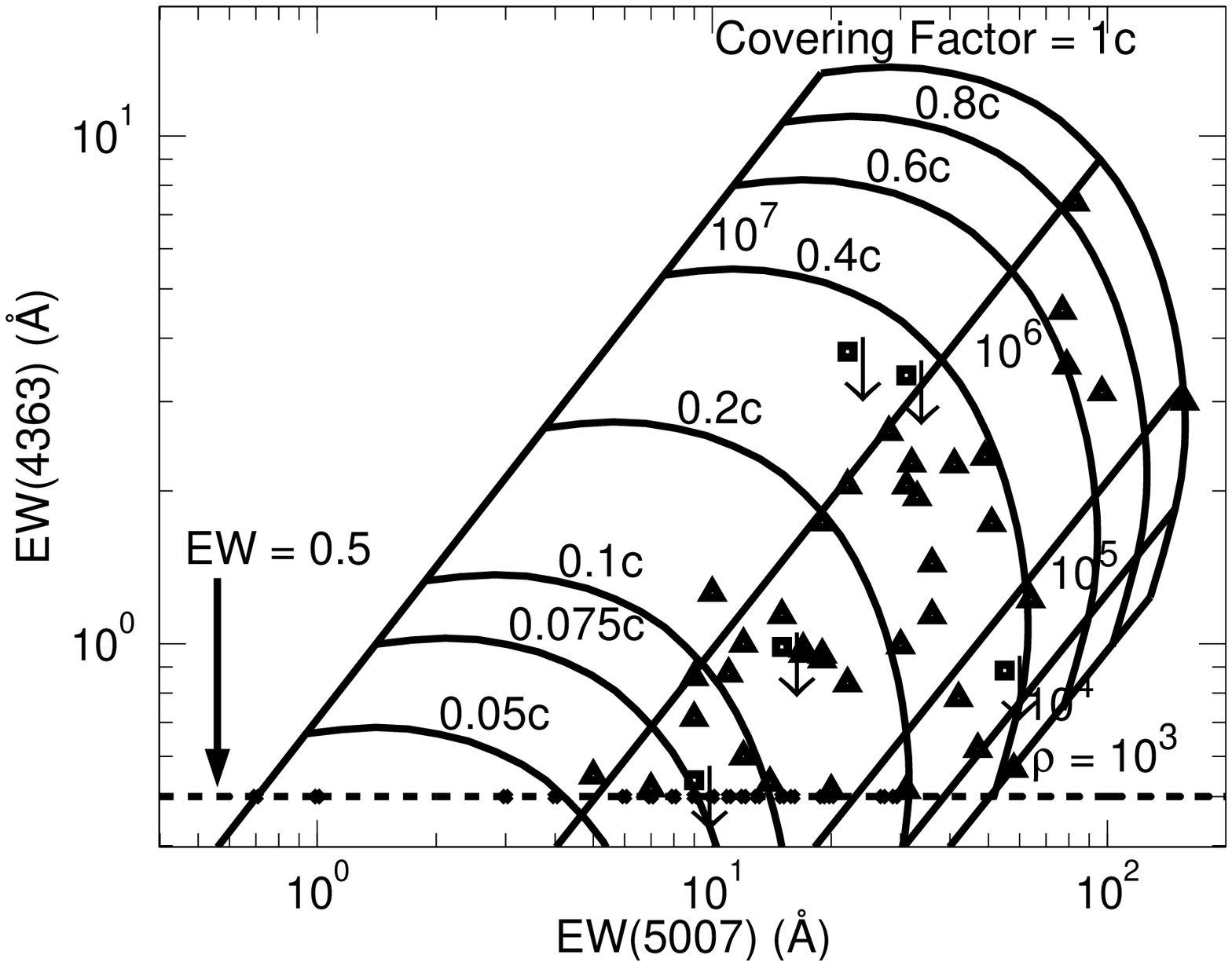}{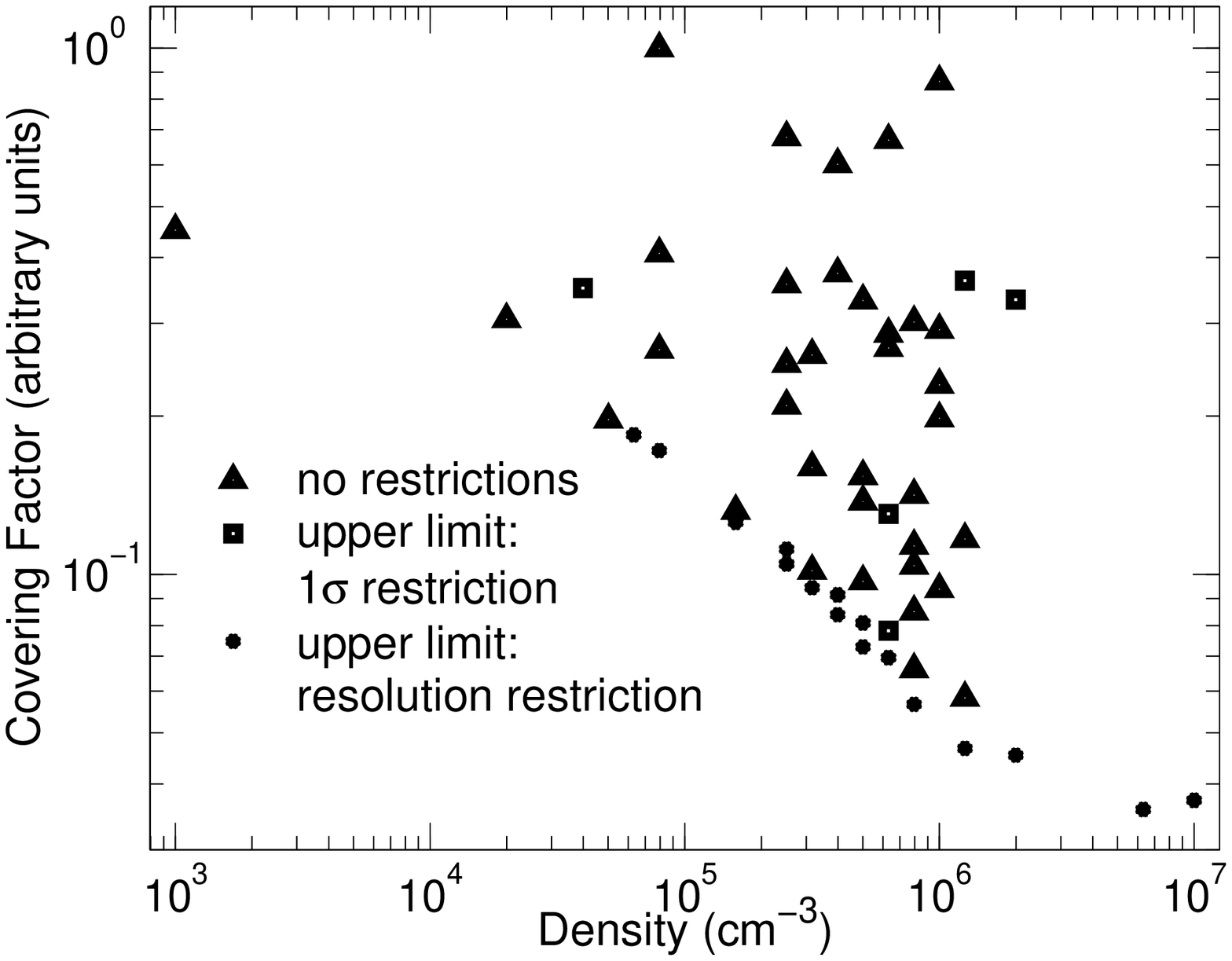} \caption{Left: the
theoretical relations (solid curves) vs. the observed positions
of the 72 BQS quasars with a measureable [O~III]$\lambda 5007$ and
[O~III]$\lambda 4363$ lines.
Note that EW(4363) = 0.5~\AA\ is our
adopted lower detection limit. Right: the covering factor vs. the inferred
density. Note that the two parameters display a comparable spread.}
\end{figure}

\section{The [O~III] EW}

To determine whether the near absence of narrow line emission,
in particular of the [O~III]$\lambda 5007$ line, is due to NLR
densities above critical, or to a low NLR covering factor, we measure
the amount of [O~III]~$\lambda 4363$ emission from the same gas
which emits the O~III]$\lambda 5007$ line. The critical density
of [O~III]~$\lambda 4363$ is $3\times 10^7$~cm$^{-3}$ while the critical density
of  [O~III]$\lambda 5007$ is $7\times 10^5$~cm$^{-3}$. Thus, the
[O~III]~$\lambda 4363$/[O~III]$\lambda 5007$ flux ratio should
increase with increasing density in the above range. If the
NLR is suppressed by a low covering factor then
both [O~III] lines should vary together.

We measure the [O~III]~$\lambda 4363$ flux using the
[O~III]$\lambda 5007$ profile as a template for the 72 objects where
the 4363~\AA\ region was observed and where [O~III]$\lambda 5007$ is
measureable. Thus, we
measure the [O~III]~$\lambda 4363$ line from gas which has the same
velocity distribution as that of the [O~III]$\lambda 5007$ emitting gas. The
measurements are plotted in Figure 3 (left panel) together with the
theoretical curves of
the [O~III]~$\lambda 5007$ EW vs. the [O~III]~$\lambda 4363$ EW
as a function of density as calculated using CLOUDY
(assuming $\log U = -2$ and $\Sigma =10^{23}$~cm$^{-2}$ which optimizes the
[O~III] emissivity). These calculations allow us to infer the density and
relative covering factor of each object (Fig.3, right panel).
Both the density and covering factor appear to have a significant
effect on the large range in [O~III]$\lambda 5007$ EW.

Note that the absence of objects at low density and low covering
factor is due to the [O ~III]~$\lambda 4363$ detection limit,
while the lack of object at $n>10^6$~cm$^{-3}$ most likely results
from the range of NLR densities coupled with the [O~III]$\lambda 5007$
line emissivity drop above the critical density.

\section{Conclusions}

We find the following: 1. The C~IV EW correlation with luminosity (the Baldwin
relation) appears to be driven mostly by the significantly stronger
correlation of the C~IV EW with {\Ledd}. The C~IV EW is also related
to the [O~III] EW, $\alpha_{\rm ox}$, and the Fe~II/{\Hb} flux ratio.
2. Using the C~IV FWHM instead of the {\Hb} FWHM to estimate {\Ledd}
yields only a weak correlation of {\Ledd} with the C~IV EW. This suggests that
the C~IV FWHM provides a significantly less accurate measure of
{\Ledd}, compared to the {\Hb} FWHM. 3. Both the density and the covering
factor appear to contribute significantly
to the large range in the [O~III]$~\lambda 5007$
EW in AGN.

\end{document}